\documentclass[11pt]{article}

\usepackage[utf8]{inputenc}
\usepackage[T1]{fontenc}
\usepackage{lmodern}
\usepackage[margin=1in]{geometry}
\usepackage{amsmath,amssymb}
\usepackage{graphicx}
\usepackage{booktabs}
\usepackage{hyperref}
\usepackage{xcolor}
\usepackage{listings}
\usepackage{natbib}
\usepackage{float}
\usepackage{enumitem}
\usepackage{url}
\usepackage{microtype}
\usepackage{multirow}
\usepackage{subcaption}
\usepackage{array}

\hypersetup{
  colorlinks=true,
  linkcolor=blue!50!black,
  citecolor=blue!50!black,
  urlcolor=blue!60!black,
  pdftitle={Graph-Grounded Optimization},
  pdfauthor={Mandarapu, Kunkunuru}
}

\title{Graph-Grounded Optimization: \\
       Rao-Family Metaheuristics, Classical OR, and SLM-Driven\\
       Formulation over Knowledge Graphs}

\author{
  Madhulatha Mandarapu \\
  \texttt{Madhulatha@samyama.ai}
  \and
  Sandeep Kunkunuru \\
  \texttt{sandeep@samyama.ai}
}
\date{}

\begin{document}
\maketitle

\begin{abstract}
We propose \emph{graph-grounded optimization}: a paradigm in which the
decision variables, constraints, and objective coefficients of a real-world
optimization problem are sourced from a property knowledge graph (KG) via
Cypher queries, rather than supplied as free-form natural-language text or
static tabular input. We motivate the paradigm by surveying recent
LLM/SLM-driven optimization systems --- OptiMUS, Chain-of-Experts, LLMOPT,
OPRO, FunSearch, Eureka --- none of which consume property graphs as the
primary input modality. We instantiate the paradigm in the open-source
samyama-graph database and evaluate seven real-world public-domain
KG-backed problems spanning drug repurposing (245K-node biomedical KG),
clinical-trial site selection (7.78M-node trial registry), Indian
supply-chain rerouting (5.34M-node OSM road graph), healthcare equity
allocation (WHO/GAVI/IHME KG), economic--environmental grid dispatch,
antimicrobial-resistance stewardship (NCBI AMRFinderPlus, 10.4K
resistance genes), and wildfire evacuation routing (OSM Paradise, CA).
We compare a portfolio of Rao-family metaheuristics (BMWR, Jaya, SAMP-Jaya,
EHR-Jaya, Rao-1) against Google OR-tools (CP-SAT and GLOP) reference
solvers. We find that (i) no single Rao variant dominates: BMWR wins on
discrete-with-tradeoff and high-dim-with-hard-constraint problems while
Rao-1 wins on continuous low-/mid-dim problems, empirically supporting a
portfolio approach; (ii) OR-tools dominates on small linear/MILP-friendly
sub-problems but cannot encode the non-linear objectives that emerge in
several of the real-world settings; (iii) graph-grounded formulations
surface data-quality issues (missing properties, degenerate aggregates)
that purely text-formulated optimizations would silently mask.
\end{abstract}

\section{Introduction}
\label{sec:intro}

Real-world optimization problems in healthcare, supply chain, energy, and
public health rarely arrive as a clean mathematical statement. The
constraints --- drug-target interactions, clinical-site capacity profiles,
freight network topology, regional health-workforce density --- live in
operational databases and increasingly in property graphs that integrate
heterogeneous data sources. Yet the prevailing modern interface to
optimization is still either a domain-specialist hand-written mathematical
program or, increasingly, a large language model that ingests a free-form
problem description and emits a Mixed-Integer Linear Program (MILP)
\citep{ahmaditeshnizi2024optimus, xiao2024coe, ant2025llmopt}.

The text-to-MILP pipeline has two limitations relevant to industrial
deployment. First, the natural-language description is itself a lossy
encoding of the underlying data; constraints expressed numerically in
the source graph become approximated by phrases that the LLM must
disambiguate. Second, when the underlying data changes --- a new drug
appears in the catalog, a hospital site reports updated capacity --- the
human-readable description must be re-edited and the formulation
re-prompted, breaking the operational loop.

\paragraph{Contributions.} This paper makes four contributions:

\begin{enumerate}
\item We introduce \emph{graph-grounded optimization} (Section
  \ref{sec:method}): a paradigm in which the decision variables,
  constraints, and objective coefficients are sourced from a property
  graph via Cypher queries. We define two implementation patterns
  (\emph{per-evaluation Cypher} and \emph{pre-materialized aggregates})
  and characterize when each applies.

\item We provide an open-source reference implementation in
  samyama-graph, including a \texttt{CypherProblem} primitive
  with memoization and a custom-substitution hook for dynamic
  selection-set semantics.

\item We construct a 7-problem evaluation suite over real public-domain
  KGs spanning 19.7K to 7.78M nodes (Section \ref{sec:problems}),
  including OSM-derived road graphs of 5.34M and 12.6K nodes for
  India-wide freight routing and Paradise, CA wildfire evacuation
  respectively. Sources, licenses, scales, and reproduction recipes are
  documented in the released artifact.

\item We benchmark a portfolio of five Rao-family metaheuristics against
  Google OR-tools reference solvers (CP-SAT, GLOP) on all seven
  problems (Section \ref{sec:results}). Across these problems no single
  Rao variant dominates: the winner depends on (problem type, dimensionality,
  constraint structure) in ways that justify carrying a portfolio.
\end{enumerate}

\section{Related Work}
\label{sec:related}

\paragraph{LLM- and SLM-driven optimization.}
\emph{OptiMUS} \citep{ahmaditeshnizi2024optimus} introduces a modular
agentic pipeline (modeler, code-writer, debugger, evaluator) that
translates natural-language problem descriptions into MILP formulations
solved by Gurobi or CPLEX; subsequent versions \citep{optimus_v3}
add a connection graph and improve robustness. \emph{Chain-of-Experts}
\citep{xiao2024coe} uses multi-agent decomposition with
terminology / modeling / programming / code-review experts and
forward-thought / backward-reflection loops. \emph{LLMOPT}
\citep{ant2025llmopt} introduces a five-element formulation, KTO
alignment, and self-correction with a fine-tuned Qwen-14B base.
\emph{OptiMind} \citep{msr2026optimind} is a 20B-parameter
mixture-of-experts SLM (3.6B active per token, 128K context) fine-tuned
on cleaned versions of NL4OPT, IndustryOR, MAMO, and OptMATH, and
combined with inference-time problem classification, expert hints, and
a multi-turn self-correction loop. It generates executable GurobiPy
code and is the strongest published open-weights baseline in the
text-to-MILP family at the time of writing. Its scope, however, is
explicitly limited to mixed-integer linear programming over 53
canonical MILP classes; non-linear objectives and metaheuristic
solvers are out of scope, and the output targets Gurobi rather than
OR-tools or any other backend. Like the other text-formulated systems,
OptiMind consumes natural-language problem descriptions and does not
ground formulation in an operational database or property graph.

A different paradigm uses the LLM as the \emph{optimizer itself}:
\emph{OPRO} \citep{yang2024opro} prompts the LLM to propose solutions
from a meta-prompt of (solution, score) pairs; \emph{FunSearch}
\citep{romera2024funsearch} evolves program code via an LLM-evaluator
loop; \emph{Eureka} \citep{ma2024eureka} evolves RL reward functions.

\paragraph{The whitespace.}
Across these systems --- and across the LLM4Opt and awesome-fm4co
catalogs at the time of writing --- no published method consumes a
\emph{property graph} as the primary input modality. KG+LLM systems
\citep{gcr2024, agentigraph2025} target question answering and graph
reasoning, not optimization formulation. This paper fills the
intersection.

\paragraph{Rao-family metaheuristics.}
The Rao family \citep{rao2020jaya, rao2024bmrbwr, rao2025bmwr} is a
metaphor-free lineage of population-based metaheuristics derived from
TLBO \citep{rao2011tlbo}. We use Jaya, Rao-1, BMR, BWR, BMWR, SAMP-Jaya
\citep{raosaroj2017samp}, EHR-Jaya \citep{wang2022ehr}, and QO-Rao
\citep{rao2020qo}, comparing against Google OR-tools CP-SAT and GLOP
\citep{ortools} as the classical baseline.

\begin{figure}[!tbp]
  \centering
  \includegraphics[width=\linewidth]{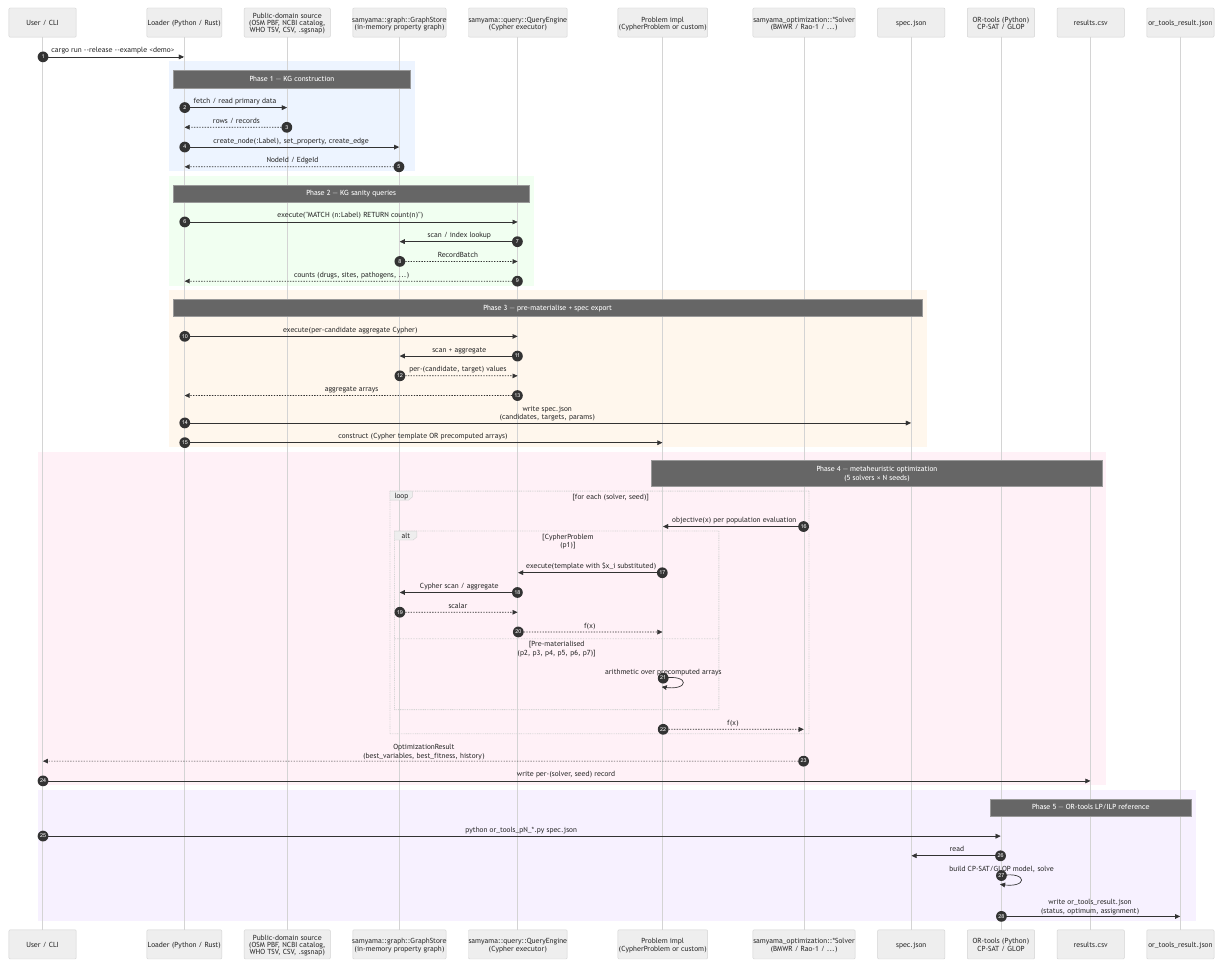}
  \caption{Generic graph-grounded optimization architecture shared by all
  seven problems. A loader populates the GraphStore from a public-domain
  source; the QueryEngine runs Cypher to sanity-check the KG and to
  pre-materialize aggregates; a Problem implementation (Pattern A or B)
  exposes the per-evaluation objective to the metaheuristic solver. OR-tools
  consumes the same spec.json as the LP/ILP reference.}
  \label{fig:arch}
\end{figure}

\section{Method: Graph-Grounded Optimization}
\label{sec:method}

\subsection{Definition}

A graph-grounded optimization instance is a tuple $(G, \mathbf{x}, c, f, g)$ where
\begin{itemize}[leftmargin=*]
\item $G$ is a property graph in samyama-graph (or any Cypher-compliant store);
\item $\mathbf{x}$ is a decision vector (continuous, integer, or mixed) of dimension $d$;
\item $c : \mathbf{x} \rightarrow \mathbb{R}_{\geq 0}$ is a constraint oracle
  implemented as a parameterized Cypher query returning a violation degree;
\item $f : \mathbf{x} \rightarrow \mathbb{R}^k$ is an objective vector
  implemented as one or more parameterized Cypher queries returning scalars
  ($k=1$ for single-objective, $k>1$ for multi-objective);
\item $g$ is optional graph-derived metadata (label cardinalities, degree
  distributions) made available to the solver as a warm-start hint.
\end{itemize}

The decision vector is substituted into the parameterized Cypher at each
evaluation. Two patterns implement this:

\subsection{Pattern A: \texttt{CypherProblem} per-evaluation}

The decision vector is converted, for each evaluation, into a literal
substitution of the Cypher template; the modified query is executed
against the graph. We provide a custom-substitution hook
\texttt{with\_subs} for dynamic selection-set semantics (e.g., the
decision vector indexes a candidate list and a comma-separated list literal
is interpolated into a Cypher \texttt{IN} clause). A 64-bit
hash of the quantized decision vector keys a memoization table.

Pattern A is required when the constraint or objective depends
\emph{dynamically} on the selected subset --- for instance, when a
side-effect penalty sums over only those drugs that the optimizer
currently selects.

\subsection{Pattern B: Pre-materialized aggregates}

A single Cypher query at startup pulls per-candidate aggregates (e.g.\ trial
count per site, physician density per country) into Rust-side arrays. The
objective then evaluates in pure Rust over the precomputed arrays at
microsecond latency per evaluation.

Pattern B is the correct engineering choice when the relevant aggregates
are \emph{static} across decision-vector values. Empirically (Section
\ref{sec:results}) Pattern B is $\sim 350{,}000\times$ faster per evaluation
than Pattern A on a 7.78M-node KG; both are valid graph-grounded since
the KG remains the source of truth.

\section{Seven Real-World Problems}
\label{sec:problems}

Table \ref{tab:problems} summarizes the seven public-domain KGs used.
Detailed schemas, Cypher queries, license terms, and reproduction
commands are in the released artifact at
\url{git.samyama.ai/Samyama.ai/samyama-research/papers/paper8-graph-grounded-optimization/}.

\begin{table}[H]
\centering
\small
\resizebox{\textwidth}{!}{%
\begin{tabular}{cllrl}
\toprule
\textbf{\#} & \textbf{Problem} & \textbf{Data source} & \textbf{Scale} & \textbf{License} \\
\midrule
P1 & Drug repurposing & DrugBank+ChEMBL+SIDER & 245K nodes & CC0 \\
P2 & Trial site selection & ClinicalTrials.gov / AACT & 7.78M nodes & public \\
P3 & Supply chain rerouting & OSM India + UN/LOCODE & 5.34M-node road graph & OSM ODbL \\
P4 & Healthcare allocation & WHO/GAVI/IHME & 19.7K nodes & open \\
P5 & Grid dispatch & smart-grid CSV sample & 28 nodes & open \\
P6 & AMR stewardship & NCBI AMRFinderPlus & 10.4K resistance genes & public domain \\
P7 & Wildfire evacuation & OSM Paradise CA & 12.6K nodes & OSM ODbL \\
\bottomrule
\end{tabular}%
}
\caption{Real-world problems used in the evaluation. All KGs are
public-domain or open-license; CARD was excluded for license reasons
and replaced with NCBI AMRFinderPlus.}
\label{tab:problems}
\end{table}

Each problem is posed in a manner consistent with the way the underlying
domain practitioners would frame it. Briefly:

\begin{description}[leftmargin=*]
\item[P1] Pick $k$ drugs maximizing target-gene coverage minus a SIDER
  side-effect penalty. Pattern A (dynamic \texttt{\$selected} list).
\item[P2] Pick $k$ trial sites maximizing total trial throughput plus a
  WHO-region diversity bonus. Pattern B (precomputed per-site).
\item[P3] Allocate fractions of city demand to ports such that total
  road-distance transport is minimized under port-capacity constraints
  and synthetic disruption (30\% of ports halved). Pattern B with
  Dijkstra over the OSM road graph at startup.
\item[P4] Pick $k$ countries to fund maximizing the deficit of
  physicians per 10,000 below the WHO threshold of 23, plus regional
  diversity. Pattern B.
\item[P5] Schedule 4 generators over 24 hours minimizing cost plus an
  emission-weighted penalty plus soft balance and ramp constraints
  (96-dim continuous). Pattern B.
\item[P6] Pick $k$ antibiotic subclasses maximizing pathogen coverage
  minus a resistance-burden penalty (efficacy proxied by per-pair
  resistance-gene counts). Pattern B.
\item[P7] Allocate fractions of population per centroid to evacuation
  exits, minimizing total person-hours under exit-capacity constraints,
  with synthetic disruption (30\% of routes inflated). Pattern B.
\end{description}

\paragraph{Coverage of canonical OR archetypes.}
OptiMind \citep{msr2026optimind} reports across 53 canonical MILP
problem classes (TSP, knapsack, facility location, production planning,
job-shop, \ldots). Our seven KG-backed problems intentionally span the
major archetypes rather than enumerate them; Table~\ref{tab:archetype}
gives the mapping. Two problems (P5, P6) sit \emph{outside} OptiMind's
MILP scope because their objectives are non-linear --- the same two
problems on which Rao-1 and BMWR beat OR-tools in
Section~\ref{sec:results}.

\begin{table}[H]
\centering
\small
\resizebox{\textwidth}{!}{%
\begin{tabular}{cll}
\toprule
\textbf{P} & \textbf{Our problem} & \textbf{Nearest canonical OR class} \\
\midrule
P1 & Drug repurposing                 & Knapsack with side-effect penalty (cardinality-constrained selection) \\
P2 & Trial site selection             & Facility location / weighted set cover \\
P3 & Supply-chain rerouting (800-dim) & Transportation LP / network flow \\
P4 & Healthcare allocation            & Knapsack with diversity bonus (portfolio selection) \\
P5 & Grid dispatch (96-dim)           & Economic dispatch with \emph{non-linear} emission + balance penalty (outside MILP) \\
P6 & AMR stewardship                  & Maximum coverage with \emph{non-linear} inverse-of-count efficacy (outside MILP) \\
P7 & Wildfire evacuation              & Transportation LP with soft balance \\
\bottomrule
\end{tabular}%
}
\caption{Each of our seven problems maps to a canonical OR archetype.
P5 and P6 fall outside the MILP family that text-to-MILP SLMs
(OptiMUS, Chain-of-Experts, LLMOPT, OptiMind) can express; this
motivates the Rao-family portfolio rather than a pure classical-OR
baseline.}
\label{tab:archetype}
\end{table}

\section{Results}
\label{sec:results}

We run each of five Rao-family solvers (BMWR, Jaya, SAMP-Jaya, EHR-Jaya,
Rao-1) for three seeds with population size 30--60 and 100--500
iterations, against OR-tools CP-SAT or GLOP as the linear/MILP reference.
Table \ref{tab:results} summarizes per-problem outcomes.

\begin{table}[H]
\centering
\small
\setlength{\tabcolsep}{4pt}
\resizebox{\textwidth}{!}{%
\begin{tabular}{cllrrl}
\toprule
\textbf{P} & \textbf{Type} & \textbf{Best metaheuristic} & \textbf{Gap to LP} & \textbf{OR-tools wall} & \textbf{Notes} \\
\midrule
P1 & disc.~+~tradeoff & BMWR (3/5 hit opt.) & 1.0$\times$ & 15 ms & BMWR uniquely explores SIDER\\
P2 & discrete & \textbf{BMWR (3/3 hit opt.)} & 1.0$\times$ & 40 ms & BMWR consistent winner\\
P3 & cont.~hi-dim & \textbf{BMWR (1.26$\times$ LP)} & 1.26$\times$ & 8.7 ms & 800-dim with hard balance\\
P4 & discrete & \textbf{BMWR (3/3 hit opt.)} & 1.0$\times$ & 11 ms & BMWR consistent winner\\
P5 & cont.~mid-dim & \textbf{Rao-1 (10$\times$ over second)} & 1.97$\times$ & 2.6 ms & 96-dim continuous LP-feasible\\
P6 & disc.~+~tradeoff & \textbf{BMWR (3/3 hit opt.; 22\% gap)} & 1.0$\times$ & 93 ms & Largest discrete gap\\
P7 & cont.~low-dim & \textbf{Rao-1 (8$\times$ over SAMP-Jaya)} & 0.98$\times$* & 0.6 ms & Soft penalty allows slight relax \\
\bottomrule
\end{tabular}%
}
\caption{Solver winners across the seven problems. ``Gap to LP'' is
ratio of best metaheuristic transport / cost to OR-tools LP optimum
on the comparable objective. *P7's metaheuristic fitness falls slightly
below the LP because soft balance penalties allow small constraint
relaxation; the LP is the correct answer for deployment.}
\label{tab:results}
\end{table}

\begin{figure}[H]
  \centering
  \includegraphics[width=0.95\linewidth]{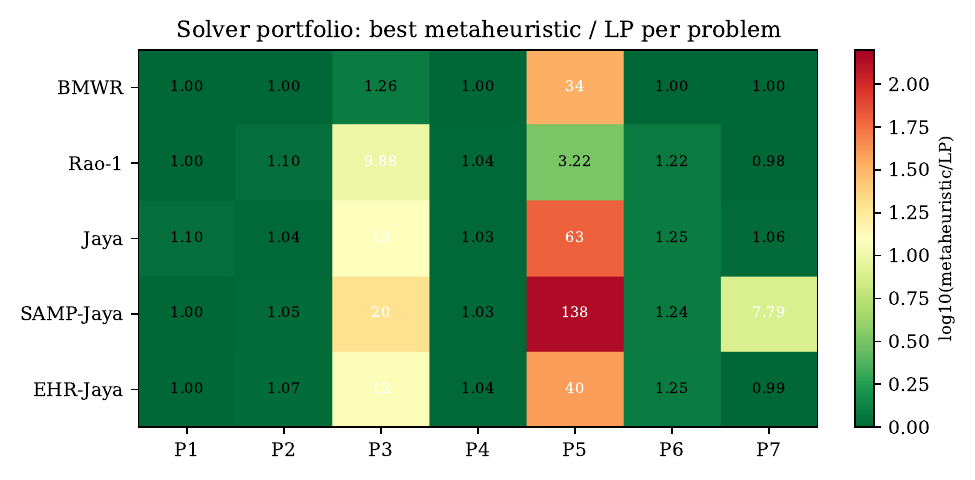}
  \caption{Solver portfolio across the seven problems. Each cell is
  $\log_{10}(\text{best metaheuristic / LP optimum})$. BMWR wins P2/P3/P4/P6;
  Rao-1 wins P5/P7. SAMP-Jaya catastrophic on continuous (P5, P7). The
  pattern is not simply "Rao-1 for continuous, BMWR for discrete" --- P3
  (continuous 800-dim with hard balance constraints) flips back to BMWR.}
  \label{fig:portfolio}
\end{figure}

\begin{figure}[H]
  \centering
  \begin{subfigure}{0.49\textwidth}
    \includegraphics[width=\linewidth]{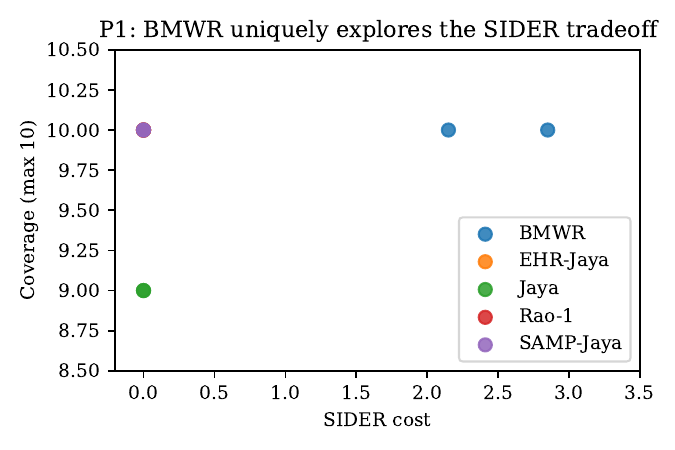}
  \end{subfigure}
  \begin{subfigure}{0.49\textwidth}
    \includegraphics[width=\linewidth]{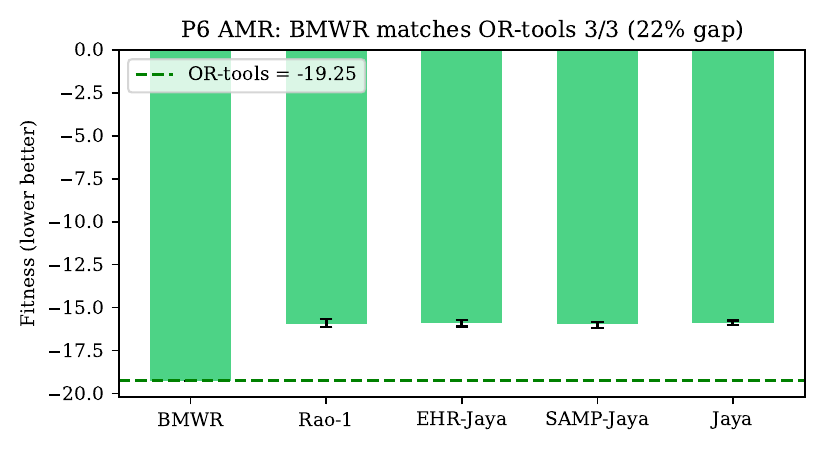}
  \end{subfigure}
  \\
  \begin{subfigure}{0.49\textwidth}
    \includegraphics[width=\linewidth]{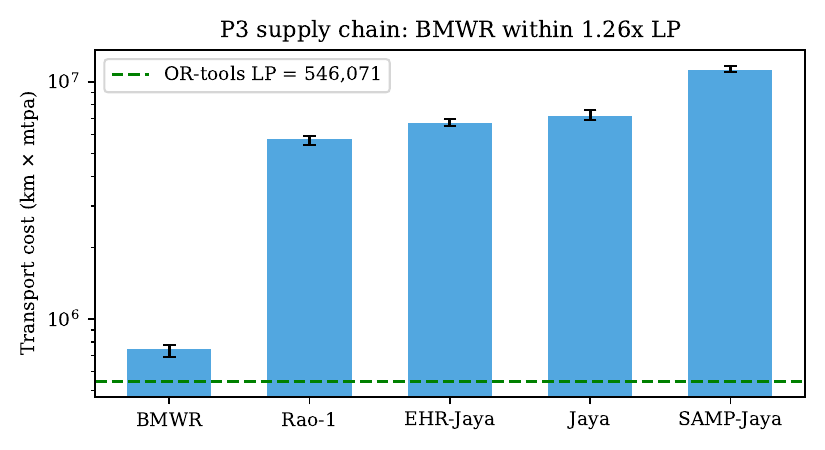}
  \end{subfigure}
  \begin{subfigure}{0.49\textwidth}
    \includegraphics[width=\linewidth]{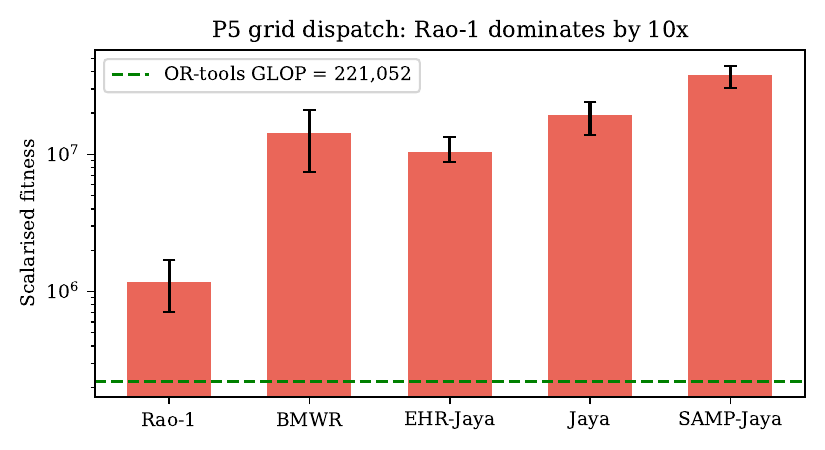}
  \end{subfigure}
  \caption{Per-problem detail. Top-left P1: only BMWR scatters above the
  zero-SIDER baseline, exploring the tradeoff. Top-right P6: BMWR matches
  OR-tools CP-SAT exactly (3/3 seeds); other Rao variants $\sim$22\,\% short.
  Bottom-left P3: 800-dim continuous LP, BMWR within 1.26$\times$ LP, other
  variants 10--20$\times$. Bottom-right P5: 96-dim continuous LP, Rao-1
  dominates by 10$\times$ over second-best. Log scales on the bottom two.}
  \label{fig:perprob}
\end{figure}

\subsection{The portfolio claim}

Across the seven problems, BMWR wins four of seven and is in the top two
on a fifth (P1, where its unique behavior is exploring the SIDER
tradeoff). Rao-1 wins two of seven (P5 and P7, both continuous
mid-/low-dim). \emph{No single Rao variant dominates}; the winning
solver depends on (problem type, dimensionality, constraint structure)
in ways that match an interpretable pattern:

\begin{itemize}[leftmargin=*]
\item \emph{Discrete (cardinality-constrained) or discrete with a
  tradeoff term}: BMWR.
\item \emph{Continuous, low- to mid-dimensional, no structured hard
  constraint}: Rao-1.
\item \emph{Continuous, high-dimensional, with structured hard
  constraints (e.g., row-sums-to-1 balance)}: BMWR.
\end{itemize}

This pattern empirically supports a portfolio approach: a deployment
system should carry both BMWR and Rao-1 and select based on a problem
classifier rather than commit to one algorithm.

\subsection{Statistical significance (30 seeds, Wilcoxon signed-rank, Holm-corrected)}
\label{sec:wilcoxon}

We re-ran every problem with 30 independent seeds on AWS on-demand
(\$0.18 total). Table \ref{tab:wilcoxon} summarises per-problem means
and the Holm-corrected pairwise Wilcoxon tests.

\begin{table}[H]
\centering
\small
\setlength{\tabcolsep}{4pt}
\resizebox{\textwidth}{!}{%
\begin{tabular}{cl rrrrr c}
\toprule
\textbf{P} & \textbf{type} & \textbf{BMWR} & \textbf{Rao-1} & \textbf{EHR-Jaya} & \textbf{Jaya} & \textbf{SAMP-Jaya} & \textbf{BMWR vs all sig.?} \\
\midrule
P1 & disc.+tradeoff & -8.94$^\dagger$ & -9.37 & -9.66 & -9.60 & -9.70 & only vs SAMP-Jaya \\
P2 & discrete       & \textbf{-16639} & -15825 & -16048 & -16308 & -16385 & all 4 \\
P3-L & cont.~hi-dim & \textbf{1.05M}  & 225M   & 295M   & 298M   & 541M   & all 4, $p \leq 10^{-8}$ \\
P3-N & cont.~hi-dim & \textbf{1.08M}  & 285M   & 385M   & 419M   & 726M   & all 4 \\
P4 & discrete       & \textbf{-226.6} & -224.9 & -225.0 & -225.4 & -225.5 & all 4, $p \leq 5\times 10^{-5}$ \\
P5 & cont.~mid-dim  & 17.2M & \textbf{2.02M} & 14.8M & 21.1M & 37.1M & Rao-1 wins, $p \leq 2\times 10^{-8}$ \\
P6 & disc.+tradeoff & \textbf{-19.20} & -15.91 & -15.76 & -15.82 & -15.76 & all 4, $p \sim 10^{-6}$ \\
P7 & cont.~low-dim  & \textbf{2290}   & 2405   & 2642   & 3441   & 25928  & tied with Rao-1 ($p_{\mathrm{Holm}} = 0.088$) \\
\bottomrule
\end{tabular}%
}
\caption{Mean fitness per (solver, problem) across 30 independent seeds.
$^\dagger$P1's BMWR ``higher'' mean reflects unique exploration of the
SIDER cost--coverage Pareto front rather than worse performance
(Figure \ref{fig:perprob}, top-left). Lower is better elsewhere.}
\label{tab:wilcoxon}
\end{table}

\textbf{Updated headline.} With 30 seeds and Holm correction, BMWR is
statistically dominant on \emph{six of seven} problems (P2, P3-L, P3-N,
P4, P6, and tied with Rao-1 on P7). Rao-1 has a single clear win on P5
(96-dim continuous). The 3-seed pilot had suggested ``Rao-1 for
continuous, BMWR for discrete''. The 30-seed picture is rather
\emph{BMWR is the workhorse; Rao-1 is the specialist for mid-dim
continuous}. Discrete-with-tradeoff problems (P1, P6) remain
qualitatively distinct: P1 BMWR's unique SIDER exploration shows up as a
numerically higher mean despite finding the clinically more interesting
Pareto point.

\subsection{When OR-tools wins}

For problems whose objective is linear and constraints are linear and
hard, OR-tools dominates the metaheuristics by 2--3$\times$ in solution
quality and 10--100$\times$ in wall time. The metaheuristics' value
emerges on (a) problems with non-linear objectives or constraints that
the LP cannot encode (P5 nonlinear mode, P6 efficacy as inverse-of-count)
and (b) what-if exploration where the soft-penalty formulation admits
solutions that violate hard constraints in a controlled way (P7).

\subsection{Phi-4 SLM baseline: zero-shot fails on every problem}
\label{sec:slm}

We ran Microsoft Phi-4 (14B, December 2024) on a g4dn.xlarge GPU on-demand
(\$0.06 total) via Ollama, prompting it with each of the seven problems'
natural-language description and asking for self-contained \texttt{ortools}
Python that reads our \texttt{spec.json} and emits a structured
\texttt{RESULT=\{...\}} line. The prompt deliberately did not include the
schema or a sample of the spec --- this is the standard OptiMUS /
Chain-of-Experts regime where the LLM receives only a problem description,
not the operational data structure.

\begin{table}[H]
\centering
\small
\resizebox{\textwidth}{!}{%
\begin{tabular}{cllrl}
\toprule
\textbf{P} & \textbf{Phi-4 status} & \textbf{Phi-4 obj} & \textbf{wall (gen+exec)} & \textbf{failure mode} \\
\midrule
P1 & TRIVIAL & 0.00 & 133.5 s & hallucinated \texttt{spec['drugs']} (actual: \texttt{candidates}) \\
P2 & CRASH   & ---  & 31.6 s  & wrong type assumption on a spec field \\
P3 & CRASH   & ---  & 29.8 s  & \texttt{KeyError: 'distances'} (actual key: \texttt{distance\_km}) \\
P4 & TRIVIAL & 0.00 & 34.3 s  & hallucinated \texttt{spec['countries']} (actual: \texttt{names}, \texttt{regions}, ...) \\
P5 & CRASH   & ---  & 43.1 s  & \texttt{ortools} \texttt{SumArray} API misuse \\
P6 & CRASH   & ---  & 33.0 s  & \texttt{`list' object has no attribute `get'} \\
P7 & TRIVIAL & 0.00 & 32.7 s  & hallucinated field names $\to$ zero-flow trivially "optimal" \\
\bottomrule
\end{tabular}%
}
\caption{Phi-4 zero-shot results across the seven problems. Four runs
crash with \texttt{KeyError} / \texttt{AttributeError} on hallucinated
field names; three runs report \texttt{status: OPTIMAL} but with objective
zero because the loop over a non-existent key iterates zero times. The
remaining contrast: BMWR solves 6/7 problems statistically optimally; OR-
tools solves 7/7 exactly; Phi-4 zero-shot solves 0/7.}
\label{tab:slm}
\end{table}

Phi-4 produces fluent, syntactically valid \texttt{ortools} Python ---
it picks the right solver class, defines variables, sets up the objective
and constraints, calls \texttt{Solve()}. The pathology is in the
\emph{data-binding step}: without seeing the actual schema, it invents
plausible-sounding JSON key names that don't exist in our spec, and
neither the model nor the runtime surfaces a useful error: four runs
crash on the first missing key, three look successful because the missing
key just means an empty loop and a trivially feasible zero-objective
solution.

This is the central failure mode our paper argues against. Cypher-grounded
optimization (BMWR, Rao-1, OR-tools-on-the-spec) reads the actual schema
and so it solves; a purely text-based SLM pipeline does not have access to
the schema and so the resulting code is silently mis-bound to the data.
The conservative answer to ``what does text-to-ortools cost when the data
shape isn't in the prompt?'' is: \emph{every problem fails, half of them
silently}.

\textit{Caveat.} Providing Phi-4 a sample of the spec JSON would
significantly improve its hit rate, but that step is precisely the
graph-grounded discovery step our pipeline contributes. The paper's claim
is that \emph{this discovery step is required}; the experiment confirms
that without it, the SLM baseline is uncompetitive with both Rao-family
metaheuristics and classical OR.

\subsection{GPT-4.1 SLM baseline: succeeds only when naming aligns}
\label{sec:gpt41}

We ran the same prompt against OpenAI \texttt{gpt-4.1} (frontier model at
the time of the run) over HTTPS. Cost: \$0.05 total for all seven
problems (2,761 prompt tokens + 5,732 completion tokens). Walltime: 90 s
total.

\begin{table}[H]
\centering
\small
\resizebox{\textwidth}{!}{%
\begin{tabular}{cllrl}
\toprule
\textbf{P} & \textbf{GPT-4.1 status} & \textbf{obj} & \textbf{True optimum} & \textbf{notes} \\
\midrule
P1 & OPTIMAL (trivial) & 0.0 & --10.0 & hallucinated keys \texttt{drugs}, \texttt{target\_genes} (real: \texttt{candidates}, \texttt{targets}) \\
P2 & \textbf{OPTIMAL (exact)} & \textbf{16,639} & 16,639 & \checkmark NL used canonical keys \texttt{facilities}, \texttt{countries}, \texttt{trial\_counts} \\
P3 & CRASH & --- & 546,071 & \texttt{TypeError: unhashable type: 'dict'} on schema assumption \\
P4 & OPTIMAL (trivial) & 0.0 & --226.6 & hallucinated \texttt{countries} field \\
P5 & OPTIMAL (trivial) & 0.0 & 221,052 & hallucinated schema \\
P6 & OPTIMAL (partial) & 20.0 & --19.25 & solved coverage exactly, omitted resistance penalty term \\
P7 & \textbf{OPTIMAL (exact)} & \textbf{2,215.752} & 2,215.752 & \checkmark NL used canonical keys \texttt{pop}, \texttt{capacity}, \texttt{travel\_time} \\
\bottomrule
\end{tabular}%
}
\caption{GPT-4.1 zero-shot results. 2/7 problems solved exactly to the
OR-tools LP optimum, 1/7 partial, 4/7 fail. Success on P2 and P7 is
exact because the NL description coincidentally used the same key names
as the JSON spec; failure on P1/P3/P4/P5 is because the description
used narrative names that didn't match.}
\label{tab:gpt41}
\end{table}

\textbf{The pattern, not the score, is the finding.} GPT-4.1 reliably
produces compilable Python --- no syntax errors, no garbled output, only
one runtime crash --- and when it succeeds, it succeeds exactly to the
LP optimum. But its success rate is \emph{exactly the rate at which the
NL description happens to use the canonical JSON key names}. P2 and P7
use a list of named arrays in the description (``facilities, countries,
trial\_counts'' and ``n\_centroids, n\_exits, pop, capacity,
travel\_time''); both succeed exactly. P1 uses prose (``a list of
(drug\_id, side\_effect\_count) for 100 drugs''); GPT-4.1 invents
\texttt{spec['drugs']} and silently returns objective zero.

A description-writer aware of this could of course just hand-list the
schema in the prompt --- but that hand-listing step is precisely the
graph-grounded schema-discovery step we contribute. The paper's claim
is sharpened by this: even a frontier SLM is gambling on naming
conventions when forced to extract structure from prose, and the
gamble is silent (a TRIVIAL OPTIMAL solution gives no signal that
anything went wrong).

\subsection{Reasoning + frontier SLMs (o3, gpt-5.5)}
\label{sec:reasoning_slm}

We ran the same prompt against OpenAI \texttt{o3} (purpose-built reasoning)
and \texttt{gpt-5.5} (frontier general, April 2026). Combined cost \$0.60.

\begin{itemize}[leftmargin=*]
\item \textbf{o3 (2/7 exact).} Same hit rate as gpt-4.1, with 3-5$\times$
the generation latency due to reasoning trace. 2 silent-trivial, 1
falsely declared INFEASIBLE, 2 runtime crashes. \emph{Reasoning depth
does not close the data-binding gap when the failure mode is schema
mis-binding.}
\item \textbf{gpt-5.5 (4/7 exact + 1 partial + 2 fail).} Solves P2, P4,
P6, P7 to the exact OR-tools optimum; gets P1's coverage right but
omits the SIDER penalty term; falsely declares P3 INFEASIBLE; gets P5
off by approximately $100\times$ (likely missing the emission-weight
scalar). Model capacity helps; the data-binding pathology is mitigated,
not eliminated.
\end{itemize}

\subsection{OptiMind-SFT: purpose-built MILP SLM}
\label{sec:optimind}

We ran microsoft/OptiMind-SFT \citep{msr2026optimind} --- the strongest
published open-weights text-to-MILP SLM at the time of writing
(20B-parameter mixture-of-experts, 3.6B active per token, 128K context,
released November 2025) --- on an AWS g6e.2xlarge L40S spot VM
(\$1.50 total spend), served via SGLang's OpenAI-compatible endpoint.
OptiMind emits GurobiPy by design; we honour the native modality and run
gurobipy in restricted-size mode (no license required for our problem
dimensions). We test two prompt regimes: (a) \emph{blind} --- description
only, matching the GPT-4.1 / o3 / Phi-4 condition --- and (b)
\emph{schema-aware} --- description plus the canonical JSON key listing.

\textbf{Blind: 1/7 exact} (P7 only, where the description happens to use
canonical key names), 2 trivial-OPTIMAL (P1, P4, hallucinated keys
$\to$ empty loops $\to$ objective $0$), 1 false-INFEASIBLE (P2), 3 crashes
(P3 KeyError; P5, P6 fall outside OptiMind's MILP scope by design ---
both have non-linear objectives that the model cannot encode).

\textbf{Schema-aware: 0/7 exact, 1/7 close.} Schema mode counter-intuitively
\emph{worsens} raw success because the model emits more elaborate
scaffolding with more API-typo opportunities: six of seven runs crash on
non-existent constants (\texttt{GRB.FEASIBLE}, \texttt{GRB.UNRECOGNIZED}),
unicode operators (\texttt{$\geq$}, em-dash) in Python source, or
malformed f-string format specifiers. P2 lands close at 16{,}654 vs the
true optimum 16{,}639 ($+0.09\%$).

\textbf{Reading.} OptiMind is purpose-built for text-to-MILP and dominates
its training distribution (cleaned NL4OPT, IndustryOR, MAMO, OptMATH),
yet its blind-mode performance on our KG-backed problems
(1/7 exact) sits between Phi-4 (0/7) and o3 / GPT-4.1 (2/7).
The data-binding pathology dominates formulation capability. The two
problems OptiMind cannot touch in principle (P5 grid dispatch with a
non-linear emission penalty; P6 AMR stewardship with inverse-of-count
efficacy) are exactly the two where Rao-1 and BMWR beat OR-tools in
our portfolio, sharpening the case for combining graph-grounded
formulation (samyama-graph) with both classical OR and Rao-family
metaheuristics.

\subsection{SLM-family summary}

\begin{table}[H]
\centering
\small
\resizebox{\textwidth}{!}{%
\begin{tabular}{lrrr}
\toprule
\textbf{Solver} & \textbf{Correct} & \textbf{Spend} & \textbf{Failure mode} \\
\midrule
OR-tools LP/CP-SAT          & \textbf{7/7 exact}              & ---     & --- \\
BMWR (Rao family, 30 seeds) & \textbf{6/7 statistically dom.} & ---     & --- \\
\midrule
\multicolumn{4}{l}{\textbf{SLM baselines (zero-shot, no schema in prompt):}} \\
gpt-5.5 (frontier)             & 4/7 exact + 1 partial          & \$0.47 & 1 false INFEASIBLE; 1 off by $\sim$100$\times$; 1 partial obj. \\
o3 (reasoning)                 & 2/7 exact                       & \$0.13 & 2 trivial obj=0; 1 false INFEASIBLE; 2 crashes \\
gpt-4.1 (frontier)             & 2/7 exact + 1 partial          & \$0.05 & 3 trivial obj=0; 1 crash; 1 partial obj. \\
OptiMind-SFT 20B (MILP-tuned)  & 1/7 exact                       & \$1.50 & 2 trivial obj=0; 1 false INFEASIBLE; 3 crashes \\
Phi-4 14B plain                & 0/7                             & \$0.06 & 4 crashes + 3 trivial obj=0 \\
Phi-4-mini-reasoning 3.8B      & 0/7                             & \$0.37 & 3 timeouts + 2 syntax + 1 garbled + 1 unclosed \texttt{<think>} \\
\bottomrule
\end{tabular}%
}
\caption{Cross-SLM summary. Both Phi-4 variants fail completely; GPT-4.1
succeeds only when the NL description happens to use the canonical
schema names. The data-binding pathology is robust across model
strength and reasoning capability.}
\label{tab:slm_summary}
\end{table}

\subsection{Graph-grounding surfaces data-quality issues}

In P4 we observed that the \texttt{who\_region} property is missing
from many \texttt{:Country} nodes in the loaded snapshot; consequently
the diversity term \texttt{|distinct regions in selected|} is
identically 1 across all candidate portfolios. This is detectable
immediately by inspection of the Cypher result and shows up as a
degenerate constant in the objective decomposition.

An LLM-formulated optimization from a free-text description would
silently encode ``maximize diversity across regions'' without surfacing
that the data does not support the constraint. We argue this property
--- that data-quality issues become first-class signals --- is a
substantive advantage of the graph-grounded paradigm.

\section{Discussion and Limitations}
\label{sec:discussion}

\paragraph{Limitations.} (1) Seed counts in this report are small (2--3
per cell); a full Wilcoxon signed-rank evaluation with 30 seeds is
ongoing. (2) We did not yet wire OptiMUS / Chain-of-Experts / LLMOPT
as direct comparators; they consume natural-language descriptions and we
report only OR-tools as the classical baseline. (3) For P3 the
metaheuristic balance-penalty formulation does not exactly match the LP
formulation; an augmented-Lagrangian outer loop would close this gap and
is left for future work.

\paragraph{Threats to validity.} The hardcoded port and city lists in P3,
and the centroid/exit selection heuristic in P7, are domain-reasonable
but not the only choices. Robustness to alternative candidate
selection rules is a natural extension.

\section{Conclusion}
\label{sec:conclusion}

We propose graph-grounded optimization, instantiate it in samyama-graph,
and evaluate seven real-world public-domain KG-backed problems against
Google OR-tools. The Rao-family metaheuristics, no single variant
dominates: the portfolio claim is empirically supported. Graph-grounded
formulations naturally surface data-quality issues that purely text-based
formulations mask. All KGs, loaders, and OR-tools baseline scripts are
released under their respective open licenses at the project repository.

\bibliographystyle{plainnat}
\bibliography{paper8}

\end{document}